\documentclass[prl,twocolumn,superscriptaddress]{revtex4}
\usepackage{epsfig}
\usepackage{times}
\usepackage{amsmath}
\usepackage{amsfonts}
\usepackage{amssymb}
\usepackage[usenames]{color}
\usepackage{graphicx}

\newcommand{\beq}{\begin{equation}}
\newcommand{\eeq}{\end{equation}}
\newcommand{\beqa}{\begin{eqnarray}}
\newcommand{\eeqa}{\end{eqnarray}}

\newcommand{\NTT}{NTT Basic Research Laboratories, NTT Corporation, 3-1 Morinosato-Wakamiya, Atsugi, Kanagawa, 243-0198, Japan.}

\newcommand{\OSAKA}{Graduate School of Engineering Science, University of Osaka, 1-3 Machikane-yama, Toyonaka, Osaka 560-8531, Japan.}
\newcommand{\NII}{National Institute of Informatics, 2-1-2 Hitotsubashi, Chiyoda-ku, Tokyo 101-8430, Japan.}
\begin{document}
\title{Improving the coherence time of a quantum system via a coupling
with an unstable system}

\author{Yuichiro Matsuzaki}			\email{matsuzaki.yuichiro@lab.ntt.co.jp} \affiliation{\NTT}
\author{Xiaobo Zhu}		\affiliation{\NTT}
\author{Kosuke  Kakuyanagi}			\affiliation{\NTT}
\author{Hiraku  Toida}			\affiliation{\NTT}
\author{Takaaki Shimo-Oka}			\affiliation{\OSAKA}
\author{Norikazu Mizuochi}			\affiliation{\OSAKA}
\author{ Kae Nemoto}				\affiliation{\NII}
\author{Kouichi Semba}				\affiliation{\NII}\affiliation{\NTT}
\author{William J.  Munro}				\affiliation{\NTT}
\author{Hiroshi Yamaguchi} 	\affiliation{\NTT}
\author{Shiro Saito} 	\affiliation{\NTT}
\begin{abstract}
 Here, we propose a counter-intuitive use of a hybrid system where the coherence time of a
 quantum system is actually
 improved via a coupling with an unstable system. If we couple a
 \textcolor{black}{two-level system} with a single NV$^-$ center, then a dark state of the NV$^-$ center
 naturally forms after the hybridization. We show that this dark state
 becomes robust against environmental fluctuations due to the coupling
 even when the coherence time of the \textcolor{black}{two-level system} is much shorter than that of the
 NV$^-$ center.
 Our proposal opens a new way to use a quantum hybrid system for the
 realization of robust quantum information processing.
\end{abstract}

\maketitle
\textcolor{black}{
One of the promising candidates for the realization of quantum information
processing are nitrogen-vacancy (NV$^-$) center in diamond
\cite{dutt2007quantum,wrachtrup2001quantum,maze2008nanoscale, taylor2008high, balasubramanian2008nanoscale,Go01a,gruber1997scanning,jelezko2002single}.
NV defects in diamond consist of a nitrogen atom
 and a vacancy in the adjacent site, which provides a spin triplet
 system \cite{davies1976optical}.
High controllability of NV$^-$ centers
has been achieved with the current technology, \textcolor{black}{including
reliable single qubit operations that have been performed using
microwave pulses \cite{JGPGW01a}.}
Also, quantum non-demolition measurements \textcolor{black}{have been performed} by using an optical transition
 between its electron spin triplet ground state and 
 a first excited spin triplet state \cite{robledo2011high}.
Moreover, entanglement generation between distant NV$^-$ centers with the help
 of a flying photon has been reported \cite{bernien2013heralded}.
 All these properties are prerequisite
for the realization of quantum information processing.}

\textcolor{black}{
Nevertheless, NV$^-$ center is affected by dephasing due to magnetic-field environmental
noise, which limits the coherence time of the quantum states. The
dephasing time is \textcolor{black}{of the order of hundreds of microseconds using Ramsey measurements}
\cite{fang2013high, balasubramanian2009ultralong,mizuochi2009coherence, maurer2012room}.
}
\textcolor{black}{To improve the coherence time, one typically uses an active pulse control such as a spin
echo or dynamical decoupling so that the coherence time can be improved
by one or two orders of magnitude \cite{de2010universal}. These
techniques rely on the fact
that the effect of
the environmental noise is canceled out by performing single qubit
rotations at a specific timing. However, this requires a precise control
of the quantum states. Imperfection of the applied pulses accumulates as
an error, and this also degrades the fidelity of the quantum state.
}

Here, we propose a scheme to improve the coherence time of an NV$^-$ center
 in a passive way which does not require
 any active operations. Our schemes rely on a coupling the NV$^-$
 center with another two-level system (TLS). We have found that a dark state naturally forms in
an NV$^-$ center coupled with the TLS, and suggest a way to use
this dark state as a controllable qubit.
Surprisingly, even
if the TLS is much more unstable than the NV$^-$
center, this hybridization makes the coherence time of the dark state longer
than that of an NV$^-$ center alone, which is highly counterintuitive.

 \textcolor{black}{ In this paper, we espesially discuss the use of a superconductng flux qubit (FQ) as the TLS, and consider a magnetic coupling between  a FQ and an NV$^-$ center
\cite{imamouglu2009cavity,verdu2009strong,wesenberg2009quantum,schuster2010high,
wu2010storage,kubo2010strong,amsuss2011cavity, kubo2011hybrid,
zhu2011coherent,saito2013towards,marcos2010coupling,sandner2012strong,kubo2012storage} to \textcolor{black}{illustrate} our
 proposal.
We will 
consider  a gap tunable FQ (with a persistent
 current of 1 $\mu $A) that can be resonantly and strongly coupled to the NV center
 \cite{marcos2010coupling}.}
A collective coupling between a FQ and a spin ensemble of the NV centers
 has been demonstrated
 \cite{zhu2011coherent, saito2013towards}, and a coupling between a FQ
 and a single NV center has been theoretically suggested \cite{twamley2010superconducting}.
Unfortunately, the coherence time of the superconducting FQ
is not so long as the other relevant systems for quantum information
processing.
Despite many effort, the best coherence time of the superconducting FQ are on
the order of 10 $\mu$s \cite{bylander2011noise}.
However, our theoretical proposal provides a long-lived dark state of
the NV center even if the coherence time of the FQ is hundreds of
nanoseconds, which makes our protocol feasible even in the current technology.

Although we mainly discuss the FQ in this paper, we start considering a general TLS to
couple with the NV$^-$ center.
We describe the Hamiltonian of TLS, an
NV$^-$ center, and the magnetic interaction
between them as follows:
\begin{eqnarray}
 H_0&=&H_{\text{TLS}}+ H_{\text{NV}}+H_{\text{int}} \nonumber \\
 H_{\text{TLS}}&=&\frac{\hbar}{2}\epsilon \hat{\sigma }_z +\frac{\hbar}{2}\Delta \hat{\sigma
  }_x\nonumber \\
 H_{\text{NV}}&=&\hbar D\hat{S}^2_{z}+\hbar g _e\mu_B{\bf {B}}_{\text{NV}}\cdot {\bf S_{\text{NV}}}\nonumber \\
  H_{\text{int}}&=&\hbar g_e\mu _B
  \hat{\sigma }_z  {\bf {B}}_{\text{TLS}}\cdot {\bf S_{\text{NV}}}
\nonumber 
\end{eqnarray}
where $\hat{\sigma }_{x,y,z}$ denotes the Pauli operator for TLS, $\epsilon $ represents the energy
bias, $\Delta $ denotes the tunnel energy. 
\textcolor{black}{Above}, $\hat{S}_{x,y,z}$ denotes the
spin 1 operator of the NV$^-$ center, $D$ denotes a zero-field
splitting, $g _e$ denotes a g-factor, $\mu _B$ represents Bohr magneton,
${\bf {B}}_{\text{NV}}$ denotes a magnetic field applied to the NV$^-$ center,
and ${\bf{B}}_{\text{TLS}}$ denotes a magnetic field generated from the TLS.
 \textcolor{black}{It is worth
mentioning that we ignore the effect of the strain because it is
typically much smaller than the homogeneous broadening of the NV center
\cite{fang2013high}. Moreover, it is possible to cancel out the effect of the strain by
applying an electric field \cite{dolde2011electric}.}
 We can rewrite the interaction Hamiltonian as $g_e\mu _B
  \hat{\sigma }_z  {\bf {B}}_{\text{TLS}}\cdot {\bf S_{\text{NV}}}=G
  \hat{\sigma }_z \cdot
  (\hat{S}_{x,k}\cos \phi \sin \theta 
  -\hat{S}_{y,k}\sin \phi  \sin \theta +\hat{S}_z\cos \theta )$
  where \textcolor{black}{$G=g_e\mu _B|{\bf {B}}_{\text{TLS}}|$ denotes
  the coupling strength.
 \textcolor{black}{ The zero-field
spin splitting of the NV$^-$ center sets a quantization axis
to be the direction between the nitrogen
and the vacancy, which we call z axis. We define}
   $\theta $ as an angle of ${\bf{B}}_{\text{TLS}}$ from this z axis.}
Since the Hamiltonian has a
  symmetry around the z-axis, \textcolor{black}{we choose a specific direction orthogonal to
  the z axis as the x axis so that the magnetic-field angle $\phi $ from the x
  axis can be set
  as } $\phi =0$ without loss of
  generality. Also, as long as $D\gg g_e\mu_B |{\bf {B}}_{\text{NV}}|$,
  the x and y component of the magnetic field does not change the
  quantized axis of the NV center \textcolor{black}{significantly}, and so we consider only z axis of the
  field.
  \textcolor{black}{We now rotate the TLS by an angle of}
  $\cos \xi
=\frac{\epsilon }{\sqrt{\epsilon ^2+\Delta ^2}}$
\textcolor{black}{to diagonalize it so that we obtain $\hbar \frac{\sqrt{\epsilon ^2 +\Delta ^2} }{2}\hat{\sigma
    }_z$
where we use the same Pauli matrix after diagonalization}, and move to a rotating
  frame defined by $U=e^{-i(\frac{1}{2}\omega
\hat{\sigma }_z+\omega  \hat{S}_z^2)t}$
  \textcolor{black}{\cite{marcos2010coupling}. In this case,}  we can rewrite our
  Hamiltonian as
  \begin{eqnarray}
   H_0&=&\hbar \frac{\sqrt{\epsilon ^2 +\Delta ^2} -\omega }{2}\hat{\sigma
    }_z
    +\hbar G_{\perp } \cdot (\hat{\sigma}
_+\hat{\overline{S}}_{-} +\hat{\sigma }_-\hat{\overline{S}}_{+})
\nonumber \\
&+&\hbar G_{\|} \hat{\sigma }_z\hat{S}_z
 +\hbar (D-\omega )\hat{S}^2_{z}
+\hbar g _e\mu _BB_{z,\text{NV}} \hat{S}_{z}
  \nonumber 
\end{eqnarray}
where 
 $G_{\|}=G\cos
\theta \cos \xi$ and $G_{\perp}=G\sin \theta 
\sin \xi $. 
 We define $\hat{\overline{S}}_{+}=|\mathcal{B}\rangle \langle
 0|$ ($\hat{\overline{S}}_{-}=|0\rangle \langle
 \mathcal{B}|$) as a raising (lowering) operator of the NV $^-$ center
 for the bright state $|\mathcal{B}\rangle = \frac{1}{\sqrt{2}}(|+1\rangle +|-1\rangle )$ and
dark state $|\mathcal{D}\rangle =\frac{1}{\sqrt{2}}(|+1\rangle -|-1\rangle )$. 
 The
bright state is directly coupled with the TLS while there is
no coupling between the TLS and the
dark state.

\begin{figure}[bth]
\includegraphics[scale=0.3]{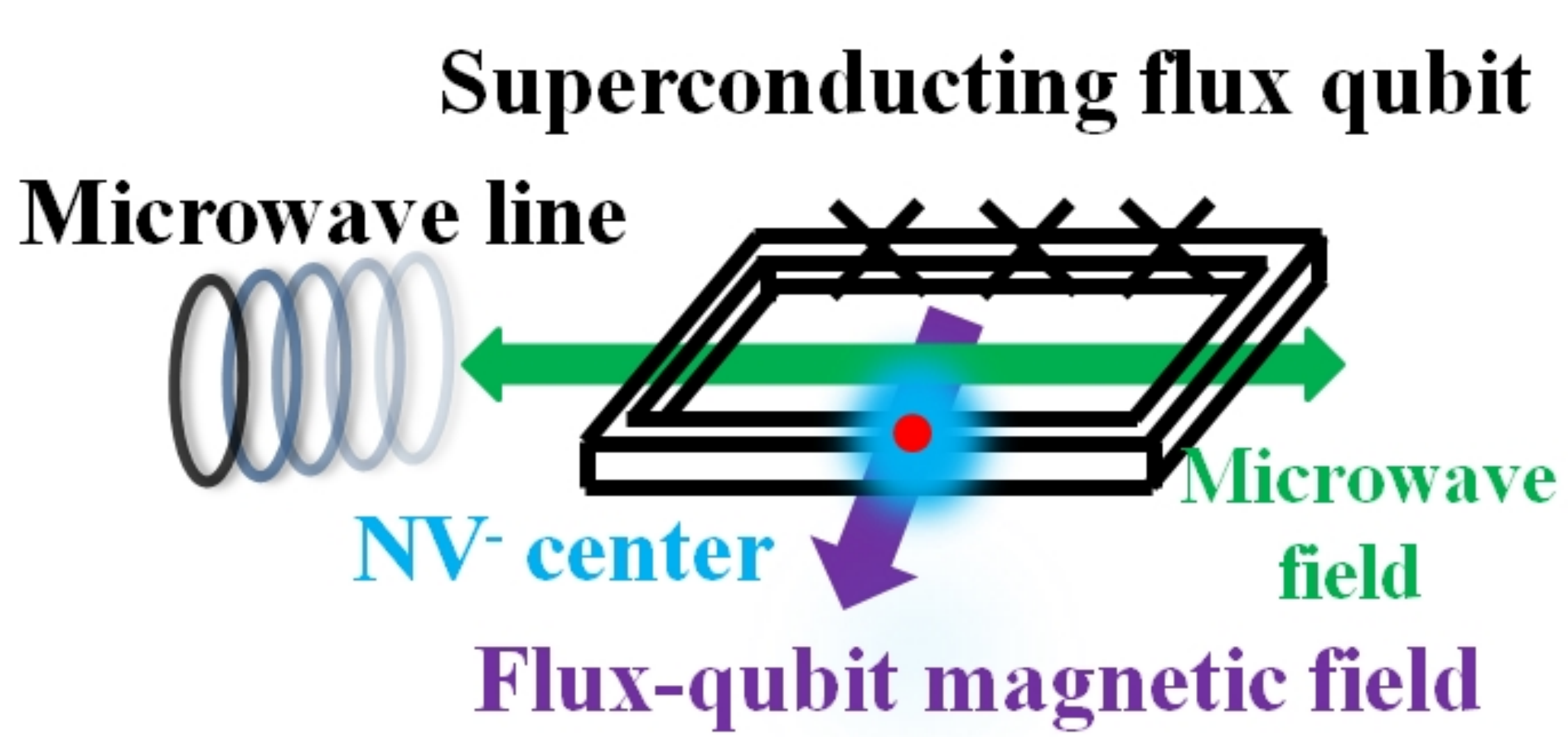}
\caption{Hybrid system composed of a superconducting flux qubit (FQ) and
 an NV$^-$ center. Since the NV$^-$ has a three-level system, \textcolor{black}{we can
 naturally} define a
 bright state and a dark state. The bright state is hybridized with the
 state of the FQ
 while the dark state has no direct coupling with the FQ. By using a
 polarization of the microwave, we can directly drive the dark state of
 the NV$^-$ center.
 }
 \label{schematic}%
 \end{figure}

Among many candidates to have a magnetic coupling with
an NV$^-$ center, we especially discuss a
superconducting flux qubit (FQ) to couple an NV$^-$ center
\cite{zhu2011coherent,saito2013towards,marcos2010coupling} as described in the FIG. \ref{schematic}.
The FQ has an
advantage to tune the frequency
\cite{zhu2010coherent, fedorov2010strong} and also has a
strong coupling due to a persistent current around 1 $\mu $A \cite{paauw2009tuning}, which
makes the FQ as an attractive system to realize our proposal.
 \begin{figure}[bth]
\includegraphics[scale=0.156]{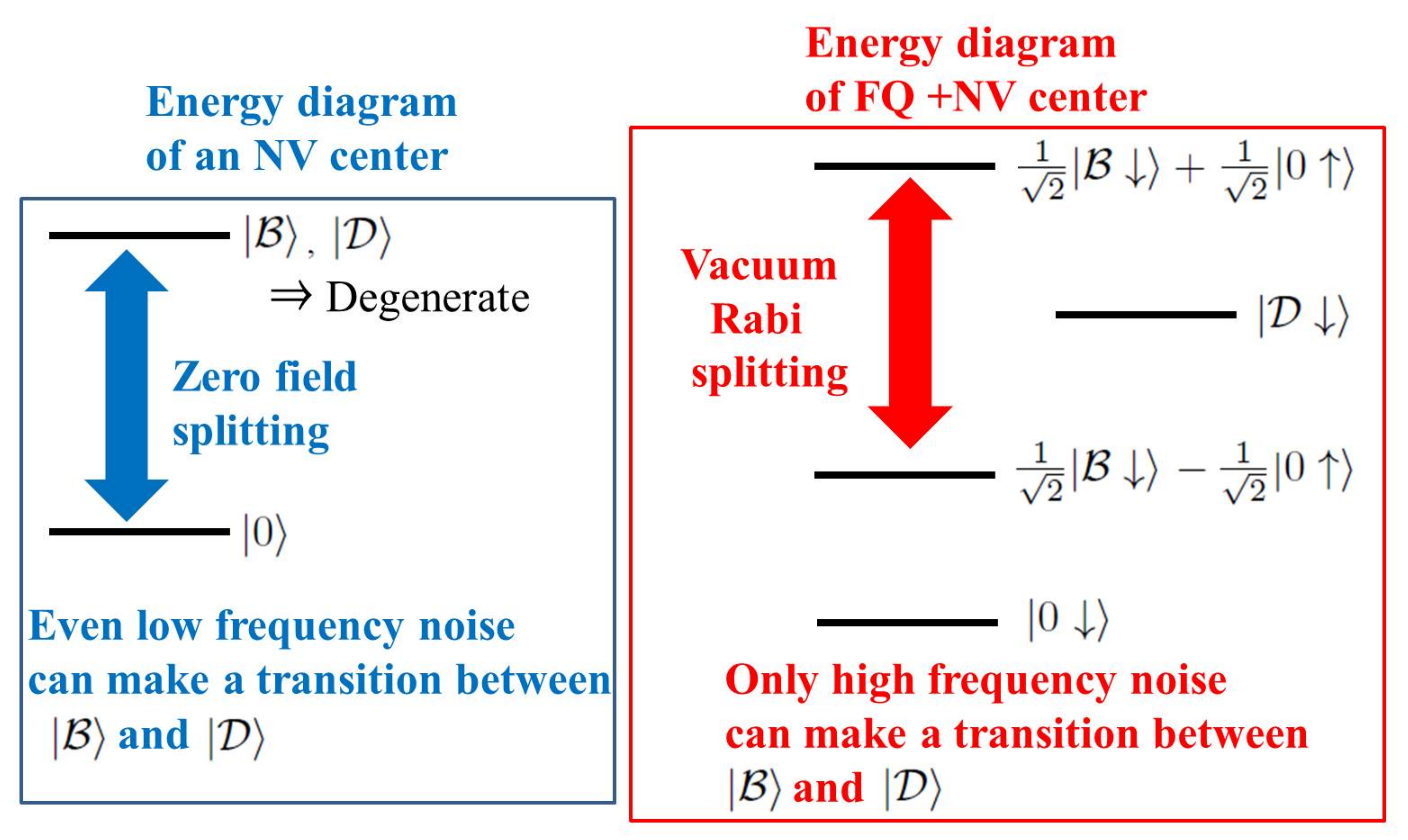}
\caption{Energy diagram of a single NV$^-$ center and a hybrid system
  composed of an NV$^-$ center and superconducting flux qubit (FQ). For a single
  NV$^-$ center, low frequency fluctuations of a magnetic field can induce an
  unwanted transition between $|\mathcal{B}\rangle $ and $|\mathcal{D}\rangle $, which
  decreases the coherence of the states. However, once the NV$^- $
  center is coupled with the FQ, low frequency component for the magnetic noise cannot induce such transition due to the
  energy splitting of the hybrid system. This mechanism makes
  the coherence time of the states much longer than that of a single NV$^-$ center
  alone.}
 \label{diagram}%
 \end{figure}

When we set $\Delta  =D$ and $\epsilon =0$ by tuning
the parameters of the FQ and set $g_e\mu _BB_{z,
\text{NV}}=G_{\|}$ by controlling the applied external magnetic field,
we can diagonalize this Hamiltonian, and also the coherence time of the FQ
becomes maximized for $\epsilon =0$ \cite{KakuyanagiMenoSaitoNakanoSembaTakayanagiDeppeShnirman01a,
YoshiharaHarrabiNiskanenNakamura01a}. We assume these conditions throughout
this paper.
The eigenvalues are given as
$E_1=-\frac{D-\omega }{2} $, $E_2=\frac{D-\omega }{2}-G_{\perp}$,
$E_3=\frac{D-\omega }{2}$, and $E_4=\frac{D-\omega }{2}+G_{\perp}$
where we consider only zero
or one excitation subspace. 
The corresponding eigenvectors in the subspace are
$|E_1\rangle =|0\downarrow\rangle $, $|E_2\rangle =\frac{1}{\sqrt{2}}| \mathcal{B}\downarrow\rangle
    -\frac{1}{\sqrt{2}}| 0\uparrow\rangle$, $|E_3\rangle =|
    \mathcal{D}\downarrow\rangle$,
and $|E_4\rangle =\frac{1}{\sqrt{2}}| \mathcal{B}\downarrow \rangle
    +\frac{1}{\sqrt{2}}| 0\uparrow\rangle$ as described in the
    FIG. \ref{diagram}.
    Since the other eigenstates $|+1\uparrow
\rangle $ and $|-1\uparrow \rangle $ that contain
two excitations are energetically
detuned,
we do not consider them.
    
    Interestingly, when we make a superposition
    $\alpha |E_1\rangle +\beta |E_3\rangle $,
    this state is robust against static (or low frequency)
    magnetic field noise.
This noise typically induces two decoherence, dephasing due to
    an energy fluctuation and unknown
    transitions to another state. Firstly, the Hamiltonian dynamics provides phase
    information with each state, and so we obtain \textcolor{black}{$\alpha|E_1\rangle +\beta e^{-\frac{i(E_3-E_1)t}{\hbar}}|E_3\rangle $}.
    If the eigenenergy of each state is \textcolor{black}{fluctuated by noise}, we have
    unknown phase shift $\theta _t$ such as \textcolor{black}{$\alpha|E_1\rangle +\beta
    e^{-\frac{i(E_3-E_1)t}{\hbar}-i\theta _t}|E_3\rangle $}, which
    causes a dephasing.  
If we add a static environmental magnetic
     field $B_{z,\text{en}}$, 
the eigenenergies $E^{(1)}_1 \simeq E_1 + \langle E_1|(g_e\mu _B
     B_{z,\text{en}} \hat{S}_z)|E_1\rangle $ and $E^{(1)}_3\simeq E_3 + \langle E_3|(g_e\mu _B
     B_{z,\text{en}} \hat{S}_z)|E_3\rangle$
       are not changed in the first-order
     perturbation theory.
    This means that
    the dephasing induced by an unknown magnetic field can be negligible for
    this state.
    Secondly, the environmental magnetic field can in principle
    induce a
    transitions between $|E_3\rangle \leftrightarrow |E_2\rangle $ and
    $|E_3\rangle \leftrightarrow |E_4\rangle $. However, these transitions require
    high frequency components of the noise to fill the energy gap
    $G_{\perp}$ between them.
    Since low-frequency magnetic field noise is the
    relevant cause of the decoherence for the NV$^-$ centers \cite{de2010universal}, such
    transitions can be also negligible due to the energy gap.
    These facts
    seem to show that, when we use $|E_1\rangle $ and $|E_3\rangle $ as
    a TLS to construct a qubit, the cooherence time of this system would be longer than that of an
    NV$^-$ center alone. To support this conjecture, we present a more detailed analysis.

We consider a decoherence
induced by the following noise Hamiltonian.
\begin{eqnarray}
 H_{\text{noise}}(t)=\frac{1}{2}\hbar g_e\mu _BB _{\text{noise}}f(t)\hat{S}_z
\end{eqnarray}
where $f(t)$ denotes a normalized classical random variable with a vanishing average
$\overline{f(t)}=0$ and $B_{\text{noise}}$ denotes an
amplitude of the 
magnetic noise from the environment \cite{foottone}.
We assume that the correlation function
$\overline{f(t)f(t')}$ depends only on $|t-t'|$.
We move into an interaction picture with respect to $H_0$.
By using the second order of the perturbative expansion and taking an ensemble average of $f(t)$, we obtain
\begin{eqnarray}
\overline{ \rho _I(t)}
 \simeq \rho _I(0)-\frac{1}{4}(\hbar g_e\mu _BB_{\text{noise}})^2\cdot \ \ \ \ \ \ \ \ \ \ \ \ \ \ \ \ \ \ \ \ \
 \ \ \
 \ \ \ \ \ \ \ \ \ \ \ \ 
 \nonumber \\
 \int_{0}^{t}\int_{0}^{t'}\overline{f(t')f(t'')}[\hat{S}_{zI}(t'), [\hat{S}_{zI}(t''),\rho _I(0)]]dt'dt''
\end{eqnarray}
where $\hat{S}_{zI}(t)=e^{\frac{iH_0t}{\hbar}}\hat{S}_ze^{-\frac{iH_0t}{\hbar}}$.
We calculate a fidelity such as
 $F=\langle \psi _0|\rho
 _I(t)|\psi _0\rangle $
where \textcolor{black}{$\rho _I(0)= |\psi _0\rangle \langle
\psi _0|$} and  $|\psi _0\rangle =\alpha |0\downarrow \rangle +\beta
|\mathcal{D}\downarrow \rangle $. 
  We obtain
  \begin{eqnarray}
       1-F   
       &\simeq &\frac{1}{4}(\hbar g_e\mu
     _BB_{\text{noise}})^2|\beta |^2t\int_{-\infty}^{\infty}d\tau \overline{f(\tau )f(0)}
     e^{-iG_{\perp}\tau }
       \nonumber 
  \end{eqnarray}
  where we assume that the correlation time of this noise is much
  shorter than the decay rate of this state. By defining \textcolor{black}{the} power spectral
  density of the noise as \textcolor{black}{$S_{\text{power}}(\nu )= \int_{-\infty}^{\infty}d\tau \overline{f(\tau )f(0)}
     e^{-i2\pi \nu \tau }$}, we obtain
  $1-F= \frac{1}{4}(\hbar g_e\mu
     _BB_{\text{noise}})^2|\beta |^2t \cdot
  S_{\text{power}}(\frac{G_{\perp}}{2\pi } )$. So the
  decoherence time of this state is described as $T _{\text{dark}}\simeq
  \frac{4}{(\hbar g_e\mu
     _BB_{\text{noise}})^2 |\beta |^2 \cdot S_{\text{power}}(\frac{G_{\perp}}{2\pi })}$. This 
  clearly shows that the
  component of $G_{\perp}$ in the spectral density is relevant to
  determine the coherence time of this quantum states.

  Since the relevant noise for NV$^-$ center is
  \textcolor{black}{the Ornstein-Uhlenbeck noise} (OUN)
  of environmental magnetic field \cite{de2010universal},
  we consider this noise to evaluate how much improvement we obtain when
  the NV$^-$ center is coupled with the FQ. The power spectrum for OUN
  is $S_{\text{OUN}}(f)=\frac{\tau _c}{1+(\pi f \tau _c)^2}$ where $\tau
  _c $ denotes a correlation time of this system. If we have
  $G_{\perp}\gg \frac{1}{\tau _c}$, we obtain $T _{\text{dark}}\simeq \frac{\tau
  _c G^2_{\perp}}{(\hbar g\mu
     _BB_{\text{noise}})^2 |\beta |^2  }$. Interestingly, this form is the same as when
  one performs a dynamical decoupling under the effect of OUN noise
  with a time interval of $\frac{\hbar }{G_{\perp}}$ \cite{de2010universal}.
  So we can interprete this result as a dynamical decoupling by using
  the FQ to protect the coherence from the environmental magnetic field noise. It is
  worth mentioning that, in our scheme, no active operation is
  required for this protection because the existence of the FQ itself
  suppresses the effect of noise while dynamical decoupling requires 
  an application of $\pi$ pulses with an appropriate time interval.
  If we prepare the state of $\frac{1}{\sqrt{2}}|\mathcal{D}\rangle
  +\frac{1}{\sqrt{2}}|0\rangle $ without the FQ, this
  decoheres and loses its coherence, which is called as a free induction decay (FID). The
  coherence time for the FID is
  described as $T^{(\text{NV})}_\text{FID}\simeq \frac{2}{\hbar g_e\mu _BB_{\text{noise}}}$
  \cite{de2010universal}. If we apply a spin echo, the effect of
  low
  frequency noise can be suppressed, which reveals a longer coherence time as
  $T^{(\text{NV})}_{\text{echo}}=(\frac{24\tau _c}{(\hbar g_e\mu _BB_{\text{noise}})^2})^{\frac{1}{3}}$
  \cite{klauder1962spectral}.
  We use values for NV
  centers $T^{(\text{NV})}_{\text{FID}}\simeq 62$ $ \mu $s and $T^{(\text{NV})}_{\text{echo}}\simeq 266
  \ \mu $s
  which corresponds to $g_e\mu _B B_{\text{noise}}= 0.032\ (\mu  s)^{-1}$ and
  $\tau _c =800$ $\mu$s
  \cite{de2010universal, fang2013high, balasubramanian2009ultralong}. On the other hand, if we have $G_{\perp}=2\pi \times 100$
  kHz, the coherence time is $T_{\text{dark}}=0.31$ s. \textcolor{black}{This coupling
  strength can be achieved when the FQ has the persistent current of
  1 $\mu $A, the distance between the surface of the FQ and the NV$^-$ center
  of $35$ nm, the FQ's wire width of $70$
  nm, the wire height of $25$ nm \cite{marcos2010coupling, twamley2010superconducting}.
  Therefore, it would be
  possible to make the coherence time of the NV center much longer than the
  $T^{(\text{NV})}_{\text{echo}}$ if the decoherence on the FQ is negligible.}

\begin{figure}[h]
 \includegraphics[width=8.25cm]{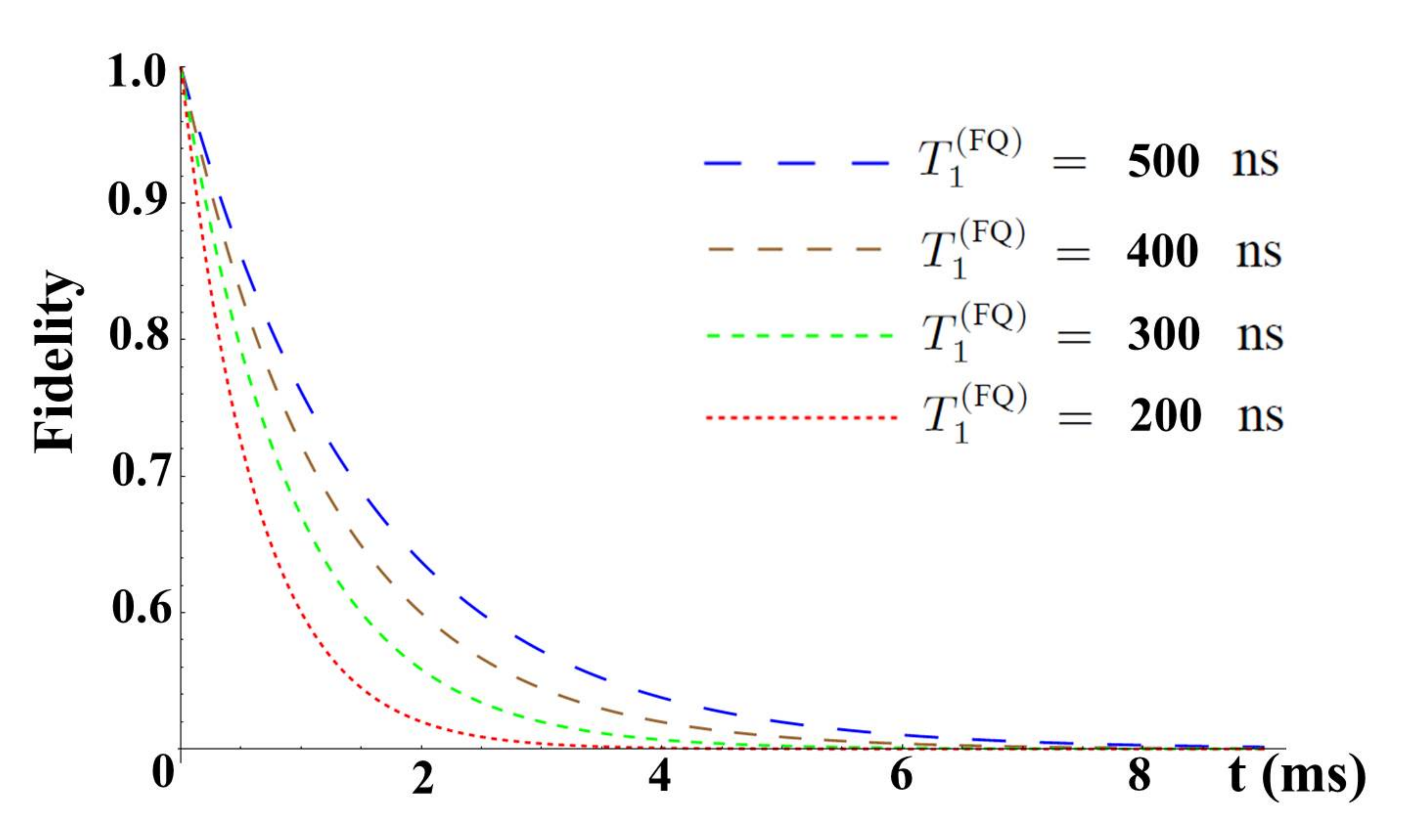}
 \caption{(color online). Fidelity of the dark state against a time.
        The initial state is $\frac{1}{\sqrt{2}}|\mathcal{D}\downarrow \rangle
    +\frac{1}{\sqrt{2}}|0\downarrow \rangle $. We assume $\tau _c =800
       $ $\mu $s, $g_e\mu _BB_{\text{noise}}=0.032$ $(\mu s)^{-1}$, $G_{\perp}=2\pi \times 100$ kHz, and $T^{(\text{FQ})}_1=200,300,400,500$ ns (from the bottom). This decay is fit by an exponential curve $\frac{1}{2}+\frac{1}{2}e^{-\Gamma t}$.
     The fitting parameter is obtained as $\frac{1}{\Gamma } =0.62,
        0.93, 1.2, 1.5$ ms.
}
 \label{33scale}
 \end{figure}

  In the discussion above, we consider only magnetic field fluctuation
  on the NV$^-$ center, and we now present a more detailed analysis to
  include other imperfection. Especially, the FQ typically has much
  shorter coherence time than the NV$^-$ center. We
  include the decoherence of the FQ, and we investigate how much
  improvement of the coherence time is possible for the NV$^-$ center in
  such a realistic circumstance.
  It is worth mentioning that the FQ is mainly affected by a white-noise energy
  relaxation \cite{bylander2011noise}.
  In order to include the energy relaxation of the FQ and \textcolor{black}{classical
  magnetic-field noise on the NV$^-$ center,}
  we use conditional state operators $\rho_j(t)$ $(j=1,2)$, and
 we average the state operators over the
 noise sample paths occupying the $j$ th state at the time $t$ \cite{PMVOJK01a}.
 The averaged state operator can be written as $\rho(t)=\sum_{j=1}^{2}
 \rho_j(t)$ where the dynamics of $\rho_j(t)$ are the solutions of the
 coupled master eqnuations as follows
   \begin{eqnarray}
   &&H_j=H_0-\frac{(-1)^j}{2}\hbar g_e\mu _BB_{\text{noise}}\hat{S}_z, \nonumber \\
 &&  \frac{d\rho _j(t)}{dt}=-\frac{i}{\hbar}[H_j,\rho
  _j(t)]-\frac{(-1)^j}{\hbar \tau _c}\Big{(}\rho
    _1(t)-\rho _2(t)\Big{)} \nonumber \\
   &-&\frac{1}{2\hbar T^{(\text{FQ})}_{1}}(\sigma _+\sigma _-
    \rho _j(t)+\rho _j(t)\sigma _+\sigma _--2 \sigma _-\rho _j(t)\sigma
    _+), \nonumber 
  \end{eqnarray}
  for $j=1,2$
 where $\rho (t)=\rho _1(t) +\rho _2(t)$, $\rho
 _1(0)=\rho _2(0)=\frac{1}{2}\rho (0)$,
 and $T^{(\text{FQ})}_1$
 denotes an energy relaxation time of the FQ. \textcolor{black}{It is worth mentioning
 that this classical magnetic-field noise has the same correltion
 function as what OUN has \cite{PMVOJK01a}.}
 We prepare a state $\frac{1}{\sqrt{2}}|\mathcal{D}\downarrow \rangle
     +\frac{1}{\sqrt{2}}|0\downarrow \rangle $, and
 we numerically investigate a decay behavior of this state under this
 master equation by plotting the fidelity $F$ as shown in
 FIG. \ref{33scale}.
 Here, except the relaxation time of the FQ,
 we use the same parameter described above. Interestingly, even when the
 coherence time of the FQ is
 hundreds of 
 ns which is much shorter than that of the NV center alone, the
 existence of the FQ makes the coherence time of the dark state longer than
 $T_{\text{NV,echo}}$.

\begin{figure}[h]
 \includegraphics[width=8.25cm]{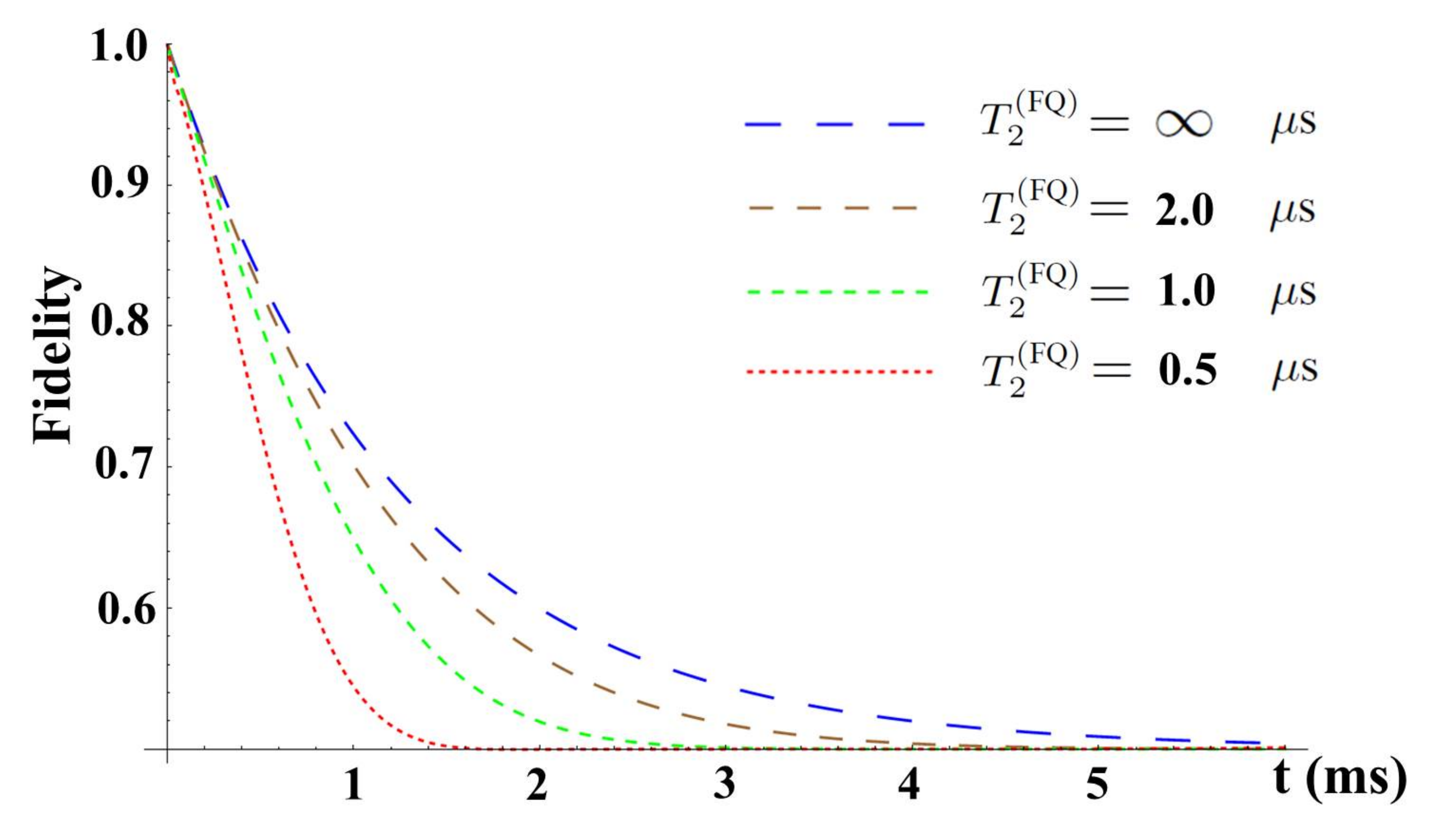}
 \caption{(color online). Fidelity of the dark state a time.
     We use $T^{(\text{FQ})}_2=0.5, 1.0, 2.0, \infty $ $\mu $s  (from the
       bottom) and $T_1=400$ ns. Other parameters are the same as those
       in FIG. \ref{33scale}.
The decay rate is obtained as $\frac{1}{\Gamma } =0.53, 0.80, 1.0, 1.2$ ms. 
}
 \label{5scale}
 \end{figure}

 Although a FQ coherence time can be in principle limited by $T_1$ process,
 it is known that there also exists
 low frequency dephasing on the FQ, and we investigate the effect of
 this type of noise for a dark state.
To consider low frequency noise for the FQ, we replace the Hamiltonian
$H_1$ and $H_2$ defined above with $H_1(f_R)=H_0+\hbar Jf_R\sigma
 _z+\frac{1}{2}\hbar g_e\mu _BB_{\text{noise}}\hat{S}_z$ and
 $H_2(f_R)=H_0+\hbar Jf_R\sigma _z-\frac{1}{2}\hbar g_e\mu
 _BB_{\text{noise}}\hat{S}_z$ where $f_R$ denotes a normalized random
 classical variable.
We define a density matrix to obey this Hamiltonian as $\rho (t,f_R)=\rho _1(t,f_R) +\rho _2(t,f_R)$, $\rho
 _1(0,f_R)=\rho _2(0,f_R)=\frac{1}{2}\rho (0)$.
 We calculate a density matrix by taking an average over all possible
 $f_R$ with Gaussian weight
 \cite{KakuyanagiMenoSaitoNakanoSembaTakayanagiDeppeShnirman01a, YoshiharaHarrabiNiskanenNakamura01a} as follows
 \begin{eqnarray}
  \rho (t)=\frac{1}{\sqrt{2\pi } }\int_{-\infty}^{\infty
   }df_Re^{-\frac{(f_R)^2}{2}}\rho (t,f_R)
 \end{eqnarray}
 Here, the dephasing time of the FQ is characterized as
 $T^{\text{(FQ)}}_2=\frac{\hbar}{\sqrt{2}J}$ when there is no relaxation
 process. We calculate the coherence time of the FQ under the
 effect of such low frequency noise where we fix $T^{(\text{FQ})}_1$
 as $400$ns and use the same parameters described above. We confirmed that, when the
 $T^{(\text{FQ})}_{2}$ is more than $500$ ns, the coherence time of the dark state is still
 much longer than $T^{(\text{NV})}_{\text{echo}}$, \textcolor{black}{as shown in
 FIG. \ref{5scale}.}

 The dark state has been intensively discussed in the cavity
 QED \cite{mucke2010electromagnetically, diniz2011strongly}. It is worth mentioning that the concept of a long-lived dark state here is completely
 different from that of the cavity QED. For example, if we confine a three-level atom composed
 of two stable ground states and an unstable excited state in a cavity, the ground states form a dark state
 with a help of a driving field and a control field \cite{mucke2010electromagnetically}.
 Since this dark state does not involves an optically active excited
 state which is much more unstable than the ground states, this dark
 state is considered to have a longer coherence time than that of other
 eigenstates.
 However, since the dark state does not change the decoherence properties of the
 ground states, the stability of this dark state is determined by an original
 coherence time of the ground states. On the other hand, in our case,
the existence of the FQ actually changes the tolerance of the NV$^-$
  center states to
 decoherence. Therefore, the coherence time of the dark state with the FQ can be longer than
 that of the original states without FQ,  
 which is completely different mechanism from that
  introduced in the previous papers \cite{mucke2010electromagnetically, diniz2011strongly}.

 We mention how to control our long-lived dark state as a
 qubit. It is necessary to implement both single qubit gates and entangling gates
 for quantum information processing.
 Firstly, we can perform a single qubit
 rotation on our qubit by using a polarization selectivity of
 the NV$^-$ center.  The FQ
 can be coupled with a bright state of the the NV$^-$ center via the
 magnetic field from the FQ.   By
 applying a microwave whose oscillating direction is orthogonal to the FQ
 magnetic field, we
 can excite only a dark state without affecting the bright state \cite{alegre2007polarization}. Secondly, it is known that, with a help of an optical
 photon, entangling operations between distant NV center can be
 constructed \cite{Cabrillo:1999p339, Bose:1999p326,Barrett:2005p363,
 bernien2012heralded}.
 So, by combining these techniques, it becomes possible to use our
 long-lived dark states as qubits for scalable quantum information
 processing.

Finally, it is worth mentioning that, although we
especially consider the FQ as a specific example to form the dark state
of the NV$^-$ center, there are many other systems to have a magnetic coupling with
an NV$^-$ center. Microwave cavity is one of the promising candidate \cite{imamouglu2009cavity,wesenberg2009quantum,schuster2010high,
wu2010storage,kubo2010strong,amsuss2011cavity, kubo2011hybrid, kubo2012storage,sandner2012strong}. Although
we have
discussed the case of a TLS, it is
straightforward to apply the calculation with the case of a harmonic oscillator.
An electron spin with a dipole-dipole interaction \cite{levitt2008spin} is
also a candidate. Hyperfine coupling with a nuclear spin can
be another candidate. We can apply the results in this paper to these
systems, and it is also possible to construct the long-lived dark state
of the NV$^-$ center.

 In conclusion, we propose a scheme to use a novel dark state of an NV$^-$
 center coupled with a superconducting flux qubit. Surprisingly, even
 when the superconducting flux qubit is much more unstable against decoherence than
 the NV center itself, the hybridization between
 the flux qubit and NV$^-$ center makes the coherence time of the dark state
 significantly longer than that of the NV$^-$ center alone. Such an
 improvement of coherence time via a coupling with an unstable system
 would open a new use of a hybrid system for the realization of
 quantum information processing.

 This work was supported in part
by JSPS through the FIRST Program, by KAKENHI(S) 25220601, and by
Commissioned Research of NICT.


\begin{thebibliography}{48}
\bibitem[{\citenamefont{foot}(2008)}]{foottone}
\bibinfo{author}{\bibfnamefont{Throughout this paper, we ignore
the thermal relaxation of the NV$^-$ center because it is known that such $T_1$
process of the NV$^-$ center is
as long as $45$ seconds at low temperature \cite{amsuss2011cavity}}}.
\end{thebibliography}

\end{document}